# Brain Tumor Segmentation by Cascaded Deep Neural Networks Using Multiple Image Scales

Zahra Sobhaninia, Safiyeh Rezaei, Nader Karimi, Ali Emami, Shadrokh Samavi
*Department of Electrical and Computer Engineering, Isfahan University of Technology, Isfahan, 84156-83111 Iran*

*Abstract*— Intracranial tumors are groups of cells that usually grow uncontrollably. One out of four cancer deaths is due to brain tumors. Early detection and evaluation of brain tumors is an essential preventive medical step that is performed by magnetic resonance imaging (MRI). Many segmentation techniques exist for this purpose. Low segmentation accuracy is the main drawback of existing methods. In this paper, we use a deep learning method to boost the accuracy of tumor segmentation in MR images. Cascade approach is used with multiple scales of images to induce both local and global views and help the network to reach higher accuracies. Our experimental results show that using multiple scales and the utilization of two cascade networks is advantageous.

## I. INTRODUCTION

Brain tumors are cells in the brain that grow abnormally. Tumors that are initiated in the brain are called primary brain tumors, while secondary tumors are initiated in other body parts such as colon, lung, and skin, etc., and then reach the brain.

Tumor types are different, based on cells they initiate from. The main types of tumors are glioma, pituitary, and meningioma. Meningiomas are prevalent in women and older adults and usually cause low-grade cancer. They generally have three grades and grow slowly. Meningiomas occur in meninges, which cover and protect the brain with three membranes.

Gliomas originate from glial cells. The role of glial cells is to support nerve cells. This type of tumor is most common in older adults. Different types of glioma are ependymomas, oligodendrogliomas, and astrocytomas. Glioma forms approximately 80% of malignant brain tumors. Pituitary tumors originate from pituitary glands, most of which are benign.

Various approaches are used for brain tumor diagnoses such as magnetic resonance imaging (MRI), CT-Scan, and ultrasound imaging. MRI is a common approach that generates pictures of brain tissues with more details [1]. Brain MR Images can be captured from three directions: axial, coronal, and sagittal [2]. Physicians use each of these directions to segment tumors and classify their types. Manual segmentation and classification of tumors are time-consuming, and human error in diagnosis should be reduced [1]. Therefore, automated segmentation and classification have been suggested for reducing human errors in diagnosis.

Nowadays, due to the spread of diseases, automatic medical image segmentation has become attractive for researchers. For example, Pereira et al. used a convolutional neural network for brain tumor segmentation [3]. Sobhaninina et al. [4] proposed a method for the segmentation of fetal head in ultrasound images. Rezaei et al. [5] presented a method for segmentation of gland in histopathology images of colon that can help pathologists to determine cancer and its degree. Since Coronary artery disease is a very common disease, Nasr-Esfahani et al [6] proposed a method that enhances X-ray angiography images and segments the vessels based on CNN. Rafiei et al. [7] presented an approach to segment the liver using 3D to 2D fully convolution networks. Jafari et al. [8] presented a method for Lesion segmentation from healthy skin to diagnose melanoma, which is a type of skin cancer. The research work of [9] presented an algorithm to enhance X-ray angiography images and detect vessel regions. Nasr-Esfahani et al. [10] proposed left ventricle segmentation in cardiac MR images by extracting the region of interest and using a CNN.

Researchers propose various methods for the segmentation of brain tumors. Some of these works use handcrafted features such as histograms of gradients and edges for brain tumor segmentation [11]. Kadkhodaei et al. segmented glioma brain tumors by using super-voxel segmentation [12]. They used neural networks for tumor segmentation and extracted saliency map and texture features like mean, variance and local binary pattern (LBP) to be used as network inputs. The output of their neural network is the segmented tumor [12].

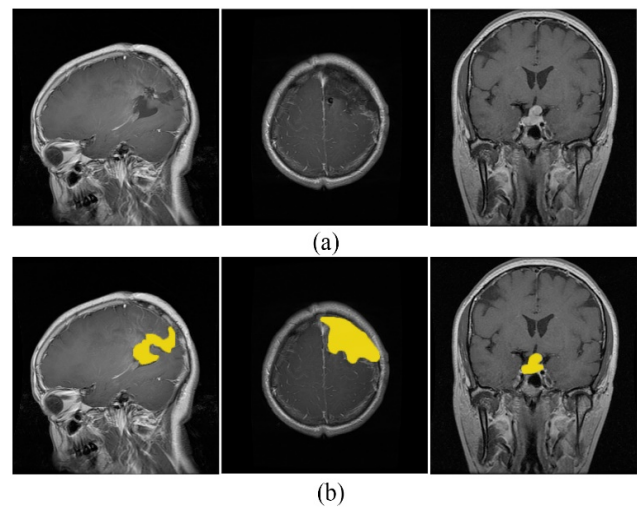

Fig. 1. Samples of Brain Tumor dataset in different directions a) Original images b) Tumor area in Ground truth provided by physicians (yellow area).

Since tumors generally comprise a small part of the image, making the images unbalanced, the use of handcrafted features cannot segment tumors well. Sobhaninia et al. [2] separated three directions of MR images and then used three convolutional neural networks to segment various types of tumors such as glioma, meningioma, and pituitary in each direction [2]. Samples of brain tumor dataset shown in Fig. 1, demonstrate that the tumor area is small in some images, and is mixed with brain tissues which make the segmentation task very challenging.

In this paper, a deep learning approach inspired by LinkNet is presented for brain tumor segmentation. In the first step, the effect of multi-scale inputs is investigated and the best approach is selected. This leads to our first contribution, the proposal of CDSL Net. In the next step, we suggest a cascaded CDSL Net which leads to significant improvement in tumor segmentation. The rest of this paper is structured as follows: Section 2 explains the proposed method in detail. Section 3 discusses the experimental results. We conclude the paper in Section 4 by a brief discussion about the proposed approach and future works.

## II. PROPOSED METHOD

### A. Overview

Figure 2 shows an overview of our proposed cascaded dual-scale LinkNet (CDSL Net). It is a multiscale network based on LinkNet, which is typically used for semantic segmentation tasks [13]. We have investigated the effects of considering other scales of the input for concatenation with corresponding layers in the network, and it is presented as a multiscale approach. Furthermore, we utilize a cascaded model framework based on DSLN for brain tumor segmentation improvement.

### B. Dual-Scale LinkNet

LinkNet is an end to end encoder-decoder architecture that is essentially a discriminative-generative network respectively [13]. As it is shown in Fig. 3 the encoder section contains parts called ResBlock, which comprise of convolutional layers with a residual link. The decoder part contains DecodeBlocks and consists of convolutional layers for the upsampling process. Some information is lost when passed from strided convolutions in the encoder part. Therefore, some skip connections from ResBlocks to corresponding DecodeBlocks are considered to help to preserve information.

One of the approaches that can be effective in retaining the image information is multi-scaling, which is feeding different scales of the input image as inputs, to be concatenated at different encoder layers. It helps to recover any lost spatial information during max-pooling and ResBlocks. We consider feeding different scales of the input image into different network layers. We experimented with different scales, such as half, one-fourth and one-eighth in each dimension. Detail results are presented in detail in section 3. There, we show that adding the half-scale has a better impact on the network's performance than employing other scales. Hence, only the half-scale image is fed as a second input. Therefore the network is called Dual-Scale LinkNet.

### C. Cascaded Dual-Scale LinkNet

Although the use of a two-scale approach increases brain tumor segmentation precision, difficulties of brain tumor segmentation like tumor location and brightness level similarity with brain tissues lead to poor results. To address these issues, the cascaded approach is suggested [14], where the second DSLN network is trained with the concatenation of the input image and output map of the previous network. The cascade approach is based on the intuition that deep convolutional networks learn a hierarchical representation of the provided data, so by cascading two networks, it is assumed that the first network learns features that are specific in the area of the brain tumor while the second network focuses more on border areas and learns more features. We show that this method leads to significantly higher segmentation quality. Quantitative evaluation results are presented in section 3.

### D. Training Cascaded-Dual-Scale LinkNet

In addition to the selection of network structure, another factor affecting the deep network's performance is choosing a loss function. Experimental results indicate that considering different loss functions affect the performance of brain tumor segmentation in MR images. In our previous work [2], the network loss function was binary cross-entropy. In this paper, the Dice criterion is also included in the loss function.

So, DSLN loss function $L_{Seg}$ is defined by (1),

$$L_{DSLN} = BCE - Dice \qquad (1)$$

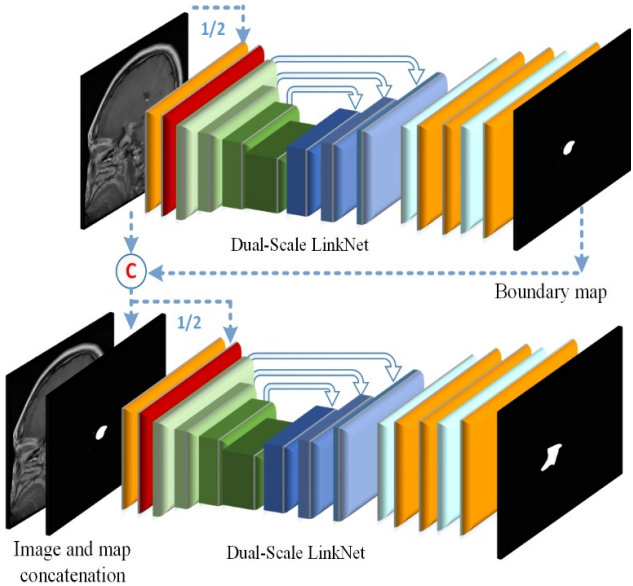

Fig. 2. Overview of the proposed cascaded Dual-Scale LinkNet

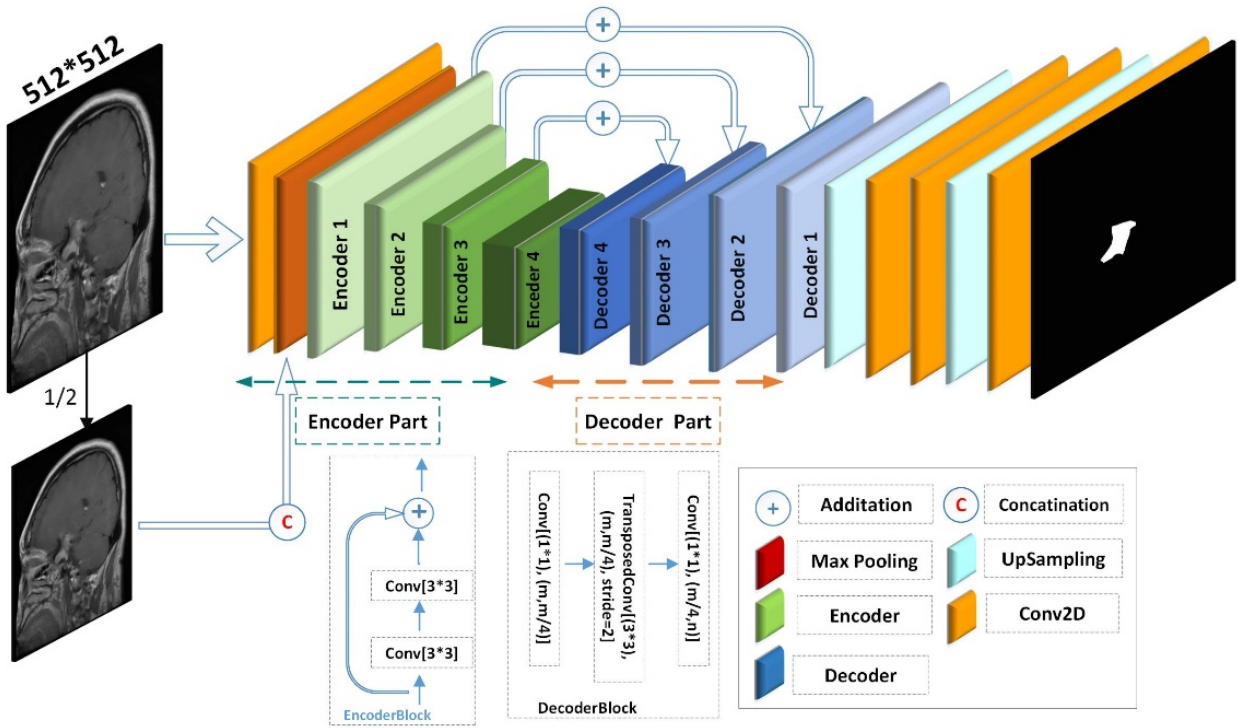

Fig. 3. LinkNet Architecture [13]

Where $BCE$ is binary cross-entropy that defined by (2), and $Dice$ coefficient is defined by (3).

$$BCE(G,I) = -(G \times \log(P) + (1-G) \times \log(1-P)) \quad (2)$$

Where $G$ is the ground-truth, $P$ is the output predicted map.

$$Dice = \frac{2\ TP}{2\ TP + FN + FP} \quad (3)$$

TP denotes true-positive, FP represents false-positive, and FN stands for false-negative.

III. EXPERIMENTAL RESULTS

The proposed network is implemented with python and tensor-flow. It is trained over 300 epochs by using SGD with momentum (learning rate = 0.001). Training time on NVIDIA GeForce GTX 1080 Ti was about 17 hours.

*A. Dataset*

In this work, the publicly available T1-weighted CE-MRI dataset[1] is used. It has 3064 grayscale images from three directions: 1047 coronal images, 1027 sagittal images, and 990 axial images. This dataset consists of three types of tumors: meningiomas, pituitary, and gliomas.

[1] http://dx.doi.org/10.6084/m9.figshare.1512427

We randomly put images into five folds to use 5-fold cross-validation. In each step, one fold is used as a test set, and other folds are used for training. Also, 20% of the training data is kept as validation data. Eventually, the average accuracy of test folds is reported as the final result.

*B. Evaluation and Quantitative results*

The evaluation score is measured with (Dice) and standard mean IoU. We initially show the effect of Loss function on the performance by comparing the proposed approach against our previous work [2].

Table 1. Comparison of Dice score with different loss function in [2]

| Loss Function | Dice |
|---|---|
| Bce [2] | 07315 |
| Bce-Dice | **0.7662** |

Results of Table 1 indicate that altering the loss function has a positive impact on the network's performance and improves the mean Dice score from 73.1% to 76.6%.
We also investigated the effect of applying different scales of the input image as extra inputs. According to the results that are shown in Table 2, Dual-Scale LinkNet has better performance on both evaluation criteria. Therefore, dual-scale is selected as the base network for brain tumor segmentation.

Table 2. Impact of Multi-Scale approach evaluation criteria

| Method | Dice | Mean IOU |
|---|---|---|
| LinkNet | 0.7662 | 0.8620 |
| Dual-Scale LinkNet | **0.7812** | **0.8813** |
| Triad-Scale LinkNet | 0.7746 | 0.8763 |
| Multi-Scale LinkNet | 0.7723 | 0.8709 |

By training a Dual-Scale LinkNet the result achieves 78% and 88% in Dice and Mean IOU, respectively. These measurements are achieved after training the second network with the output of the first network concatenated with the input image. As shown in Table 3, the cascaded model boosts both evaluation criteria.

Table 3 performance of cascaded Dual-Scale LinkNet networks on evaluation criteria

| Method | Dice | Mean IOU |
|---|---|---|
| Dual-Scale LinkNet | 0.7812 | 0.8813 |
| Cascaded Dual-Scale LinkNet | **0.8003** | **0.9074** |

*C. Evaluation and Qualitative results*

Figure 4 demonstrates the improvement of brain tumor segmentation with a cascaded approach. When the output map of the first DSL Net is concatenated with input image as input for the second DSL Net, the output result has better precision and is more similar to ground-truth.

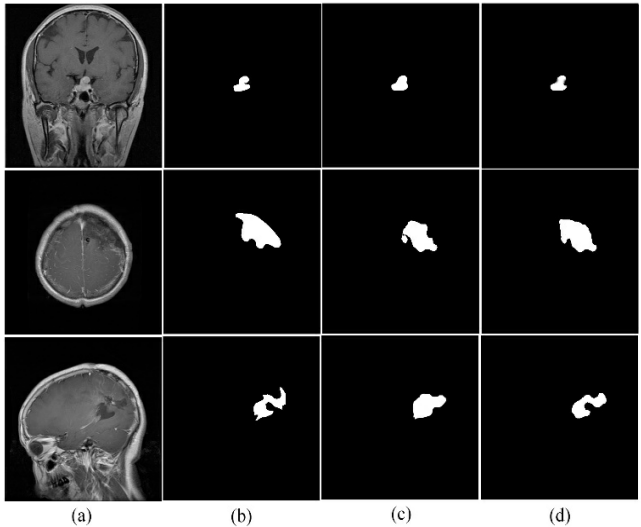

Fig. 4. Results of the brain tumor segmentation, from left to right: (a) Original image, (b) ground-truth, (c) DSL net output, and (d) CDSL Net output.

## References

[1] A. Singh Parihar , "A study on brain tumor segmentation using convolution neural network," in *International Conference on Inventive Computing and Informatics (ICICI)*, Coimbatore, India, 2017.

[2] Z. Sobhaninia, S. Rezaei, A. Noroozi, M. Ahmadi, H. Zarrabi, N. Karimi, A. Emami and S. Samavi, "Brain Tumor Segmentation Using Deep Learning by Type Specific Sorting of Images," 2018.

[3] S. Pereira, A. Pinto, V. Alves and C. A. Silva, "Brain Tumor Segmentation Using Convolutional Neural Networks in MRI Images," *IEEE Transactions on Medical Imaging,* vol. 35, no. 5, pp. 1240 - 1251, 2016.

[4] Z. Sobhaninia, S. Rafiei, A. Emami, N. Karimi, K. Najarian, S. Samavi and S. Soroushmehr, "Fetal Ultrasound Image Segmentation for Measuring Biometric Parameters Using Multi-Task Deep Learning," in *41st Annual International Conference of the IEEE Engineering in Medicine and Biology Society (EMBC)*, Berlin, Germany, Germany, 2019.

[5] S. Rezaei, A. Emami, N. Karimi and S. Samavi, "Gland Segmentation in Histopathological Images by Deep Neural Network," in *International CSI Computer Conference*, 2019.

[6] E. Nasr-Esfahani, N. Karimi, M. Jafari , S. Soroushmehr, S. Samavi, B. Nallamothu and K. Najarian, "Segmentation of vessels in angiograms using convolutional neural networks," *Biomedical Signal Processing and Control,* pp. 240-251, 2018.

[7] S. Rafiei, E. Nasr-Esfahani, K. Najarian, N. Karimi, S. Samavi and S. R. Soroushmehr, "Liver Segmentation in CT Images Using Three Dimensional to Two Dimensional Fully Convolutional Network," in *25th IEEE International Conference on Image Processing (ICIP)*, Athens, Greece, 2018.

[8] M. Jafari, N. Karimi, E. Nasr-Esfahani, S. Samavi, S. Soroushmehr, K. Ward and K. Najarian, "Skin lesion segmentation in clinical images using deep learning," in *23rd International conference on pattern recognition (ICPR)*, 2016.

[9] H. R. Fazlali, N. Karimi, S. M. R. Soroushmehr, S. Sinha, S. Samavi, B. Nallamothu and K. Najarian, "Vessel region detection in coronary X-ray angiograms," in *IEEE International Conference on Image Processing (ICIP)*, 2015.

[10] M. Nasr-Esfahani, M. Mohrekesh, M. Akbari, S. M. R. Soroushmehr, E. Nasr-Esfahani, N. Karimi, S. Samavi and K. Najarian, "Left Ventricle Segmentation in Cardiac MR Images Using Fully Convolutional Network," in *40th Annual International Conference of the IEEE Engineering in Medicine and Biology Society (EMBC)*, Honolulu, HI, USA, 2018.

[11] B. H. Menze, A. Jakab and S. Bauer, et al, "The Multimodal Brain Tumor Image Segmentation Benchmark (BRATS)," *IEEE Transactions on Medical Imaging,* vol. 34, no. 10, pp. 1993 - 2024, 2014.

[12] M. Kadkhodaei, S. Samavi, N. Karimi, H. Mohaghegh, S. M. R. Soroushmehr, K. Ward, A. All and K. Najarían, "Automatic segmentation of multimodal brain tumor images based on classification of super-voxels," in *Annual International Conference of the IEEE Engineering in Medicine and Biology Society (EMBC)*, Orlando, FL, USA, 2016.

[13] A. Chaurasia and E. Culurciello, "LinkNet: Exploiting Encoder Representations for Efficient Semantic Segmentation," in *IEEE Visual Communications and Image Processing (VCIP)*, St. Petersburg, FL, USA, 2017.

[14] L. Wu, Y. Xin, S. Li, T. Wang, P.-A. Heng and D. Ni, "Cascaded Fully Convolutional Networks for automatic prenatal ultrasound image segmentation," in *IEEE 14th International Symposium on Biomedical Imaging (ISBI 2017)*, VIC, Australia, 2017.